\documentstyle[psfig]{l-aa}

\begin{document}

\thesaurus {06 (08.02.4, 08.05.1, 08.06.3, 10.11.1, 10.15.1, 10.15.2 (Bochum~2))}

\title{Kinematics and binaries in young stellar aggregates. \\
I. \ The trapezium system BD+00$^\circ$1617 in Bochum~2\thanks{
Tables 2 and 3 are only available in electronic form at the CDS via anonymous ftp
to cdsarc.u-strasbg.fr (130.79.128.5) or via
http://cdsweb.u-strasbg.fr/Abstract.html}
}

\author{
       Ulisse Munari\inst{1}
\and   Lina Tomasella\inst{2}
       }
\offprints{U.Munari}

\institute {
Osservatorio Astronomico di Padova, Sede di Asiago, 
I-36012 Asiago (VI), Italy \ \ \ ({\tt munari@pd.astro.it})
\and
Osservatorio Astrofisico del Dip. di Astronomia, 
Universit\`a di Padova, I-36012 Asiago (VI), Italy \ \ \ ({\tt tomasella@pd.astro.it})
}
\date{Received date..............; accepted date................}

\maketitle

\markboth{U.Munari and L.Tomasella: The trapezium system BD+00$^\circ$1617
in Bochum~2}{U.Munari and L.Tomasella: The trapezium system BD+00$^\circ$1617
in Bochum~2}

\begin{abstract}
Internal kinematics, spectroscopic binaries and galactic motion are
investigated for the trapezium system BD+00$^\circ$1617 (which lies at the
heart of the young open cluster Bochum~2) by means on 73 high resolution
Echelle+CCD spectra secured over the period 1994-98. Two of the three O-type
member stars are found to be binaries on close and highly eccentric orbits
of 6.8 and 11.0 day period. The spectra of the two binaries show large
variations in the half-intensity and equivalent widths of the HeI absorption
lines, which are intrinsic to the primaries and are not correlated to the
orbital phase.  Astrometric and radial velocities exclude that one of the
component is leaving as a runaway star, but upper limits are still
compatible with the trapezium evaporating at very low relative velocities. 
The projected rotational velocities of the three constituent O--type stars
are low. This conforms to expectations from the high frequency of binaries.
The observed radial velocity of Bochum~2 agrees with the Hron (1987) expression 
for the Galaxy rotation inside the latter's quoted errors.

\end{abstract}
\maketitle

\section{Introduction}

With this paper we begin a series devoted to the results of a prolonged high
resolution spectroscopic monitoring of early type members of young open
clusters, trapezium systems and OB associations. The emphasis is on accurate
and extensive radial velocities (secured over several years), suitable to
trace ($a$) the internal kinematics, ($b$) the galactic motion, ($c$) the
binary star content and the spectroscopic orbits. The high dispersion
spectra are also useful to derive the projected rotational velocities, the
chemical composition and the evolutionary status of the member stars as well
as to study the interstellar medium inside and towards such young stellar
aggregates. These latter aspects are however broadly outside the scope of
the present series and will be considered elsewhere.

Our program encompasses a selection of young stellar aggregates spanning a
range in total mass, number of member stars, age, total energy, etc. The
targets have been chosen, whenever possible, among those with already
published high quality astrometry and photometry. We believe that much can
be learned by extensive and homogeneous high--res spectral campaign data on
such targets and to this aim we begun in 1990 a long term program with the
telescopes and spectrographs of the Asiago Astrophysical Observatory. The
ambitious goal is to search for {\sl observationally} proved grand-relations
governing the dynamical evolution of the young stellar aggregates which
could then be compared with the predictions of N-body simulations. Such
grand-relations are expected to become apparent when a significant number of
young stellar aggregates will have been studied in the most homogeneous way
and for such a reason we will postpone later in the present series a
throughout discussion of the results accumulated on the way.

We start with a so far poorly studied trapezium system (BD+00$^\circ$1617)
at the heart of a dim and very young open cluster (Bochum~2), which lies at
a great distance in the general anti--center galactic direction (at $\sim$
14 kpc galacto-centric distance).

Trapezium systems are groups of three or more stars whose apparent
separations onto the sky are similar, as for the prototype of the class, the
four O stars at the core of the Orion open cluster and nebula. OB
associations contain typically three to five trapezia and these
configurations have a high frequency among T-associations too. Within
stellar associations, trapezia appear to cluster close to the nuclei of open
clusters and emission nebulae.  A trapezium-like configuration is in general
dynamically unstable and therefore it must be young. It has to evolve either
into ($a$) {\sl hierarchical systems} (i.e. ones in which the separation of
successive binaries increases by large factors; they are dynamically stable
and are present among systems of all ages), or into ($b$) {\sl hard
binaries} (with possible ejection from the system of one or more of the
trapezium components). Ambartsumian (1954) computed in $1 \div 2 \times
10^4$ AU the largest stable dimension of a trapezium in the solar vicinity
against perturbation by passing stars: the one order of magnitude larger
extension (1$\times 10^5$ AU) for cataloged trapezia reinforces the notion
that they must be dynamically young systems. The first catalog of trapezium
systems has been compiled by Ambartsumian (1954), listing 108 entries and a
more recent one (442 systems) has been presented by Salukvadze (1978).
Trapezia have been traditionally found by filtering catalogs of visual
binaries, like the Index Catalog of Visual Binary Stars (IDS) by Jeffers et
al. (1963).

\begin{table*}
\caption[]{Program stars. Identification numbers, photometry and spectroscopy
from Munari \& Carraro (1995; MC95), positions from the Astrographic Catalog
2000 reduced to the Hipparcos Catalogue reference system.}
\begin{flushleft}
\begin{tabular}{cccccccccc} \hline
MC95 & BD        & LS & GSC     & $V$   
& {\sl U-B} & {\sl B-V} & spectrum & $\alpha_{J2000}$&$\delta_{J2000}$\\
\#&&& \# & &&&&\multicolumn{2}{c}{(epoch 1901.0)}\\
\multicolumn{10}{c}{}\\
20 & ~~+00$^\circ$1617 C~~ & ~VI +00 26~ & ~1481994~ & 11.21 & --0.52 & 0.49 & O9~V &
 06$^h$ 48$^m$ 51.$^s$300 & +00$^\circ$ 22$^\prime$ 21.$^{\prime\prime}$49 \\ 
21 &   +00$^\circ$1617 B   &  VI +00 25  &  1482125  & 11.01 & --0.48 & 0.53 & O7~V &
 06$^h$ 48$^m$ 50.$^s$484 & +00$^\circ$ 22$^\prime$ 37.$^{\prime\prime}$62 \\ 
23 &   +00$^\circ$1617 A   &  VI +00 24  &  1481965  & 11.38 & --0.52 & 0.51 & O9~V &
 06$^h$ 48$^m$ 49.$^s$541 & +00$^\circ$ 22$^\prime$ 52.$^{\prime\prime}$71 \\ 
\hline
\end{tabular}
\end{flushleft}
\end{table*}

Bochum~2 ($\alpha_{2000}=06^h 49^m, \ \delta_{2000}= +00^\circ 23^\prime, \
l = 212.3, \ b = -0.4$) has been studied by Moffat \& Vogt (1975), Turbide
\& Moffat (1993) and Munari \& Carraro (1995, hereafter MC95). At its center
lies BD+00$^\circ 1617${\sl A-B-C}, a system of three O-type stars which
luminosity largely dominates the apparently dim surrounding cluster. The
three O stars are plotted in the finding chart of Figure~1. They appear
equally spaced and on a straight line in the projection onto the sky.

Bochum~2 is not included in catalogs of trapezium systems most probably
because BD+00$^\circ 1617${\sl A-B-C} is oddly missing from catalogs of
visual binaries. However, it fully satisfies all classification criteria and
in the following it will be discussed as a validated trapezium system. The
three components will be identified in the following according to the
numbering by MC95 (\#20, \#21 and \#23, cf. Figure~1).

\section {Observations}

The program stars are listed in Table~1. Photometric and spectroscopic data
come from MC95. Positions are from the {\sl Astrographic Catalog} for the
epoch 1901.0 and equinox J2000.0 (see section 2.2), placed on the reference
system of the Hipparcos Catalogue. The journal of the observations is given
in Table~2.

\begin{table*}
\caption[]{Journal of observations. \ $N$ = spectrum number from the
telescope logbook; \ \# = program star identification number; \ {\sl HJD} =
heliocentric JD - 2400000; \ {\sl Res:} {\sl f} = full-resolution spectrum,
\ {\sl b} = $2\times 2$ binned spectrum; \ {\sl S/N} = signal-to-noise ratio
for the stellar continuum close to the NaI D doublet (center of Echelle
order 38), or center of order 26 for last entries.  Column {\sl Orders}
lists the Echelle spectral orders covered on the given spectrum, which
corresponds to the following wavelength ranges:
{\sl 23-29} = 9900--7600 \AA, {\sl 28-38} = 8130--5800 \AA,
{\sl 29-41} = 7850--5375 \AA, {\sl 33-51} = 6900--4320 \AA,
{\sl 33-52} = 6900--4240 \AA, {\sl 34-51} = 6695--4320 \AA, 
{\sl 34-50} = 6695--4410 \AA, {\sl 34-53} = 6695--4160 \AA,  
{\sl 37-50} = 6150--4410 \AA, {\sl 38-57} = 5990--3865 \AA, 
{\sl 39-58} = 5835--3800 \AA. The HJD of the last three spectra is the same
because their spectra have been recorded simultaneously on the same CCD
frame.}
\begin{flushleft}
\begin{tabular}{rccccccrcccccc}
\hline
N & \# & Date & HJD & Res. & Orders & S/N & N & \# & Date & HJD & Res. & Orders & S/N \\ 
\multicolumn{14}{c}{}\\
17522 & 21 & 13.11.94 & 49669.554 & f & 38--57 &29 &~~~~~~~~~23901 & 21 & 22.02.97 & 50502.365 & b & 34--53 &49\\    
17524 & 20 & 13.11.94 & 49669.584 & f & 38--57 &26 &    23903 & 20 & 22.02.97 & 50502.389 & b & 34--53 &50\\
17526 & 23 & 13.11.94 & 49669.622 & f & 38--57 &27 &    24457 & 21 & 25.03.97 & 50533.367 & b & 34--53 &21\\
17869 & 21 & 16.12.94 & 49703.411 & f & 29--41 &33 &    24459 & 20 & 25.03.97 & 50533.384 & b & 34--53 &18\\
17871 & 20 & 16.12.94 & 49703.438 & f & 29--41 &35 &    24461 & 21 & 25.03.97 & 50533.400 & b & 34--51 &17\\
17873 & 23 & 16.12.94 & 49703.472 & f & 29--41 &32 &    24477 & 21 & 27.03.97 & 50535.266 & b & 34--53 &32\\
18336 & 21 & 15.01.95 & 49733.379 & f & 34--53 &39 &    24479 & 20 & 27.03.97 & 50535.278 & b & 34--53 &35\\
18337 & 20 & 15.01.95 & 49733.410 & f & 34--53 &48 &    24483 & 21 & 27.03.97 & 50535.306 & b & 34--53 &21\\
18339 & 23 & 15.01.95 & 49733.457 & f & 34--53 &47 &    24485 & 20 & 27.03.97 & 50535.322 & b & 34--53 &32\\
21191 & 23 & 10.12.95 & 50061.650 & b & 34--53 &33 &    24487 & 23 & 27.03.97 & 50535.346 & b & 34--53 &36\\
21237 & 23 & 11.12.95 & 50062.589 & b & 34--53 &38 &    24489 & 21 & 27.03.97 & 50535.369 & b & 34--53 &28\\
21239 & 21 & 11.12.95 & 50062.602 & b & 34--53 &39 &    24491 & 20 & 27.03.97 & 50535.386 & b & 34--53 &24\\
21241 & 20 & 11.12.95 & 50062.616 & b & 34--53 &29 &    24651 & 20 & 16.04.97 & 50555.282 & b & 34--50 &16\\
21276 & 23 & 12.12.95 & 50063.558 & b & 34--53 &32 &    24653 & 21 & 16.04.97 & 50555.295 & b & 34--53 &34\\
21595 & 23 & 31.03.96 & 50174.372 & b & 34--53 &18 &    24655 & 20 & 16.04.97 & 50555.312 & b & 34--53 &31\\
21597 & 21 & 31.03.96 & 50174.389 & b & 34--53 &15 &    24699 & 20 & 17.04.97 & 50556.297 & b & 34--53 &45\\
21696 & 20 & 06.04.96 & 50180.314 & b & 34--53 &29 &    24701 & 21 & 17.04.97 & 50556.315 & b & 34--53 &49\\
21698 & 20 & 06.04.96 & 50180.334 & b & 34--53 &23 &    24703 & 23 & 17.04.97 & 50556.332 & b & 34--53 &48\\
21700 & 21 & 06.04.96 & 50180.368 & b & 34--53 &26 &    24770 & 20 & 18.04.97 & 50557.290 & b & 34--53 &39\\   
22588 & 23 & 30.12.96 & 50447.578 & b & 33--52 &41 &    24772 & 21 & 18.04.97 & 50557.313 & b & 34--53 &42\\
22590 & 21 & 30.12.96 & 50447.591 & b & 33--52 &57 &    24774 & 20 & 18.04.97 & 50557.332 & b & 34--53 &26\\
22592 & 20 & 30.12.96 & 50447.607 & b & 33--52 &35 &    26468 & 20 & 14.11.97 & 50766.603 & b & 34--53 &38\\
23061 & 20 & 27.01.97 & 50475.501 & b & 34--53 &30 &    26472 & 21 & 14.11.97 & 50766.650 & b & 34--53 &45\\
23063 & 21 & 27.01.97 & 50475.514 & b & 34--53 &34 &    26474 & 20 & 14.11.97 & 50766.673 & b & 34--53 &42\\
23065 & 23 & 27.01.97 & 50475.527 & b & 34--53 &26 &    26506 & 20 & 15.11.97 & 50767.595 & b & 34--53 &30\\
23168 & 20 & 28.01.97 & 50476.505 & b & 34--53 &23 &    26510 & 21 & 15.11.97 & 50767.637 & b & 34--53 &32\\
23170 & 21 & 28.01.97 & 50476.521 & b & 34--51 &24 &    26512 & 20 & 15.11.97 & 50767.661 & b & 34--53 &25\\
23172 & 23 & 28.01.97 & 50476.538 & b & 34--51 &19 &    26518 & 20 & 15.11.97 & 50767.700 & b & 34--53 &25\\
23644 & 20 & 17.02.97 & 50497.358 & b & 33--51 &32 &    27068 & 20 & 13.12.97 & 50795.522 & b & 28--38 &24\\
23645 & 21 & 17.02.97 & 50497.381 & b & 33--51 &36 &    27070 & 21 & 13.12.97 & 50795.546 & b & 28--38 &31\\
23647 & 23 & 17.02.97 & 50497.404 & b & 33--51 &36 &    27471 & 20 & 11.01.98 & 50825.461 & b & 39--58 &54\\
23723 & 23 & 18.02.97 & 50498.353 & f & 33--50 &23 &    27473 & 21 & 11.01.98 & 50825.485 & b & 39--58 &83\\
23725 & 21 & 18.02.97 & 50498.401 & f & 33--50 &16 &    27474 & 20 & 12.01.98 & 50825.515 & b & 39--58 &72\\
23823 & 21 & 21.02.97 & 50501.342 & f & 34--53 &38 &    27653 & 20 & 05.02.98 & 50850.450 & b & 23--29 &45\\
23825 & 20 & 21.02.97 & 50501.390 & f & 34--53 &36 &    27653 & 21 & 05.02.98 & 50850.450 & b & 23--29 &60\\
23827 & 23 & 21.02.97 & 50501.437 & f & 34--53 &29 &    27653 & 23 & 05.02.98 & 50850.450 & b & 23--29 &38\\
23899 & 23 & 22.02.97 & 50502.330 & f & 34--53 &40 &          &    &          &           &   &        &  \\
\hline
\end{tabular}
\end{flushleft}
\end{table*}

\begin{figure}
\centerline{\psfig{file=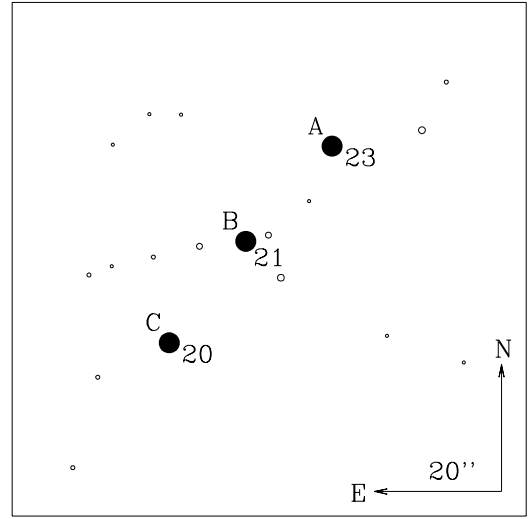,height=8cm,width=8cm}}
\caption[]{Finding chart for the program trapezium stars at the heart of the
open cluster Bochum~2. Orientation and scale are given. Capital letters
relates to BD+00$^\circ$1617 sub-classification, the numbers to the
identification in Munari \& Carraro (1995).}
\end{figure}

\begin{figure}
\centerline{\psfig{file=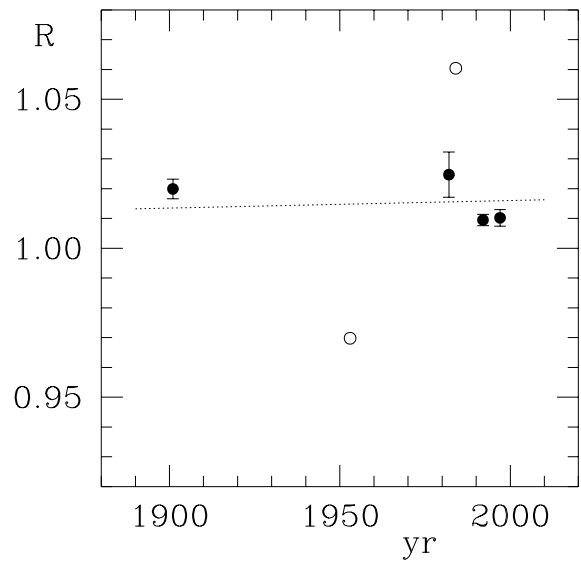,height=8cm,width=8cm}}
\caption[]{Change with time of the distance between stars $A$ and $B$
compared to that between stars $B$ and $C$ [R = $\bigtriangleup
(\overline{AB}/ \overline{BC}$)]. The dashed line is a last square fit and
the error bars are 3$\sigma$ values. Solid circles are from the Astrographic
Catalogue, the USNO Twin Astrographic Catalogue and some recent CCD
observations, the open circles from Palomar Schmidt POSS-I (USNO A1.0
Catalogue) and GSC-I.}
\end{figure}

\begin{table*}
\caption[]{Heliocentric radial velocities and associated errors for the three
program stars. \ $N^\circ$= spectrum number as in Table~2.}
\begin{flushleft}
\begin{tabular}{ccrrcccrrcccrr} \hline
\multicolumn{14}{c}{}\\                                                    
\multicolumn{4}{c}{\#20}&&\multicolumn{4}{c}{\#21}&&\multicolumn{4}{c}{\#23}\\
\multicolumn{14}{c}{}\\ \cline{1-4} \cline{6-9} \cline{11-14} 
N& HJD & RV$_\odot$ & err. &&N& HJD & RV$_\odot$ & err. &&N& HJD & RV$_\odot$ & err.\\
\multicolumn{14}{c}{}\\                                                                                    
17524 &  49669.584&   54.7 &  1.3 &&  17522 &  49669.554&   83.8 &  1.2 && 17526 &  49669.622& 68.1 & 2.9\\ 
17871 &  49703.438&   56.0 &  0.4 &&  17869 &  49703.411&   54.7 &  2.7 && 17873 &  49703.472& 68.3 & 0.3 \\ 
18337 &  49733.410&   44.0 &  1.1 &&  18336 &  49733.379&  133.7 &  1.5 && 18339 &  49733.457& 69.2 & 2.2 \\ 
21241 &  50062.616&   48.9 &  1.4 &&  21239 &  50062.602&   71.4 &  1.3 && 21191 &  50061.650& 65.5 & 3.6 \\ 
21696 &  50180.314&  101.5 &  2.9 &&  21597 &  50174.389&  135.2 &  1.5 && 21237 &  50062.589& 67.0 & 1.0 \\ 
21698 &  50180.334&  113.4 &  0.4 &&  21700 &  50180.368&    7.3 &  2.0 && 21276 &  50063.558& 68.5 & 1.1 \\ 
22592 &  50447.607&  103.9 &  1.7 &&  22590 &  50447.591&   34.2 &  2.5 && 21595 &  50174.372& 65.8 & 3.6 \\ 
23061 &  50475.501&  107.3 &  0.8 &&  23063 &  50475.514&   46.9 &  0.7 && 22588 &  50447.578& 67.6 & 1.9 \\ 
23168 &  50476.505&   89.3 &  6.0 &&  23170 &  50476.521&   47.5 &  5.4 && 23065 &  50475.527& 68.0 & 3.0 \\ 
23644 &  50497.358&   78.5 &  0.9 &&  23645 &  50497.381&   69.9 &  0.7 && 23172 &  50476.538& 70.3 & 5.5 \\ 
23825 &  50501.390&   46.7 &  1.4 &&  23725 &  50498.401&   51.3 &  1.1 && 23647 &  50497.404& 69.2 & 1.3 \\ 
23903 &  50502.389&   85.1 &  1.8 &&  23823 &  50501.342&   15.8 &  1.9 && 23723 &  50498.353& 67.7 & 2.5 \\ 
24459 &  50533.384&   47.5 &  0.9 &&  23901 &  50502.365&   22.1 &  1.1 && 23827 &  50501.437& 67.8 & 2.2 \\ 
24479 &  50535.278&   44.8 &  0.8 &&  24457 &  50533.367&    3.4 &  3.6 && 23899 &  50502.330& 71.5 & 2.7 \\ 
24485 &  50535.322&   44.3 &  3.2 &&  24461 &  50533.400&    5.2 &  3.8 && 24487 &  50535.346& 69.2 & 1.0 \\ 
24491 &  50535.386&   50.2 &  1.2 &&  24477 &  50535.266&   13.5 &  3.5 && 24703 &  50556.332& 68.8 & 2.1 \\ 
24651 &  50555.282&   50.7 &  0.4 &&  24483 &  50535.306&   20.7 &  2.5 && 27553 &  50850.450& 69.0 & 7.0 \\ 
24655 &  50555.312&   48.0 &  0.8 &&  24489 &  50535.369&    6.0 &  3.6 &&       &        &         &     \\ 
24699 &  50556.297&   51.7 &  4.9 &&  24653 &  50555.295&   13.0 &  2.2 &&       &        &         &     \\ 
24770 &  50557.290&  106.4 &  0.8 &&  24701 &  50556.315&    8.9 &  5.0 &&       &        &         &     \\ 
24774 &  50557.332&   97.3 &  0.4 &&  24772 &  50557.313&   17.9 &  1.1 &&       &        &         &     \\ 
26468 &  50766.603&   50.3 &  6.0 &&  26472 &  50766.650&   12.5 &  3.9 &&       &        &         &     \\  
26474 &  50766.673&   53.7 &  2.9 &&  26510 &  50767.637&   39.0 &  4.6 &&       &        &         &     \\
26506 &  50767.595&   52.8 &  1.1 &&  27070 &  50795.546&   51.0 &  2.4 &&       &        &         &     \\
26512 &  50767.661&   50.9 &  4.7 &&  27473 &  50825.485&  138.9 &  2.0 &&       &        &         &     \\
26518 &  50767.700&   47.7 &  3.1 &&  27553 &  50850.450&   51.9 &  5.1 &&       &        &         &     \\
27068 &  50795.522&   48.9 &  7.4 &&        &           &        &      &&       &        &         &     \\
27471 &  50825.461&  115.5 & 15.0 &&        &           &        &      &&       &        &         &     \\
27474 &  50825.515&  110.8 &  1.5 &&        &           &        &      &&       &        &         &     \\
27553 &  50850.450&   49.9 &  8.0 &&        &           &        &      &&       &        &         &     \\
\hline
\end{tabular}
\end{flushleft}
\end{table*}

\begin{table*}
\caption[]{Orbital elements for stars \#20 and \#21. The last row gives the
weighted deviation of the observed radial velocities from the computed
orbital solution. The quoted errors are the formal errors of the
orbital solution. \ {\sl HJD} = heliocentric JD -- 2400000.}
\begin{flushleft}
\begin{tabular}{lllllll} \hline
&&&\\
&&\multicolumn{2}{c}{\#20}&~~~~~~~&\multicolumn{2}{c}{\#21}\\  \cline{3-4}\cline{6-7}
&&&\\
period               & (days)          
                     & 6.858   &$\pm$0.004                
                     &&11.030  &$\pm$0.006       \\ 
baricentric velocity & (km sec$^{-1}$)~~~~~~~~ 
                     & 69.8      &$\pm$ 1.5       
                     && 63.0     & $\pm$ 1.6      \\
semi-amplitude       & (km sec$^{-1}$) 
                     &  33.3     & $\pm$ 2.3       
                     && 64.9     & $\pm$ 2.6      \\
eccentricity         &                 
                     & 0.43      &$\pm$ 0.17     
                     && 0.21     & $\pm$ 0.05     \\
$a$sin$i$            & (AU)            
                     & 0.0019    &$\pm$ 0.0001   
                     && 0.065    & $\pm$ 0.003    \\
$T_\circ$            & (HJD)           
                     & 50831.786 &$\pm$ 0.278    
                     && 50835.749& $\pm$ 0.434    \\
$\Omega$             & (deg)           
                     & 316       &$\pm$ 19       
                     && 322       &$\pm$ 14      \\
mass function        && 0.019   &
                     && 0.29     &                \\
deviation            & (km sec$^{-1}$)
                     & 4.2       &
                     && 4.8      &                \\ 
\hline
\end{tabular}
\end{flushleft}
\end{table*}

\begin{figure}
\centerline{\psfig{file=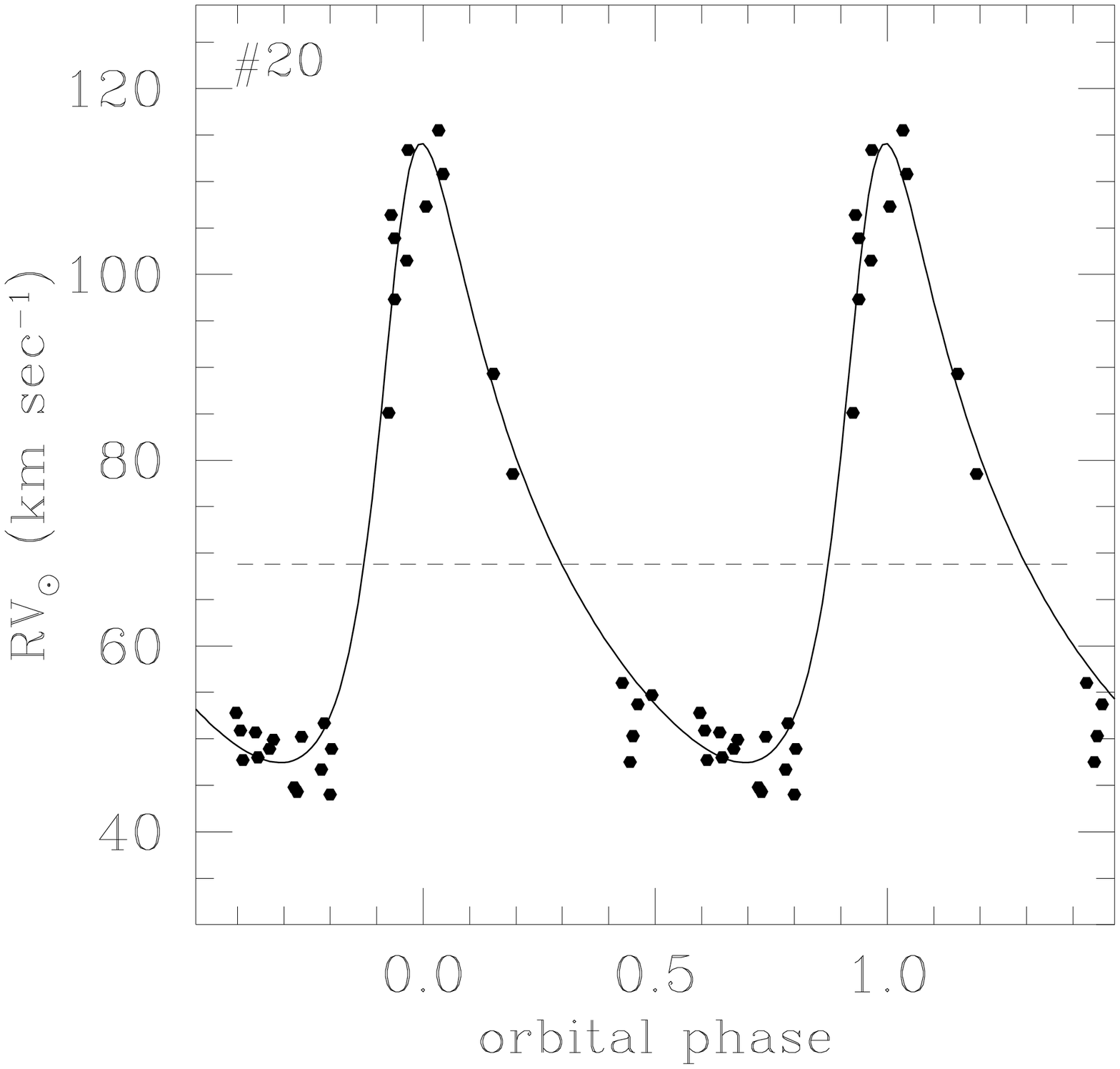,height=8cm,width=8cm}}
\caption[]{Orbital solution for star \#20. The solid line refers to the
orbital solution given in Table~4. The dashed line is the baricentric 
velocity (69.8 km sec$^{-1}$).}
\end{figure}

\begin{figure}
\centerline{\psfig{file=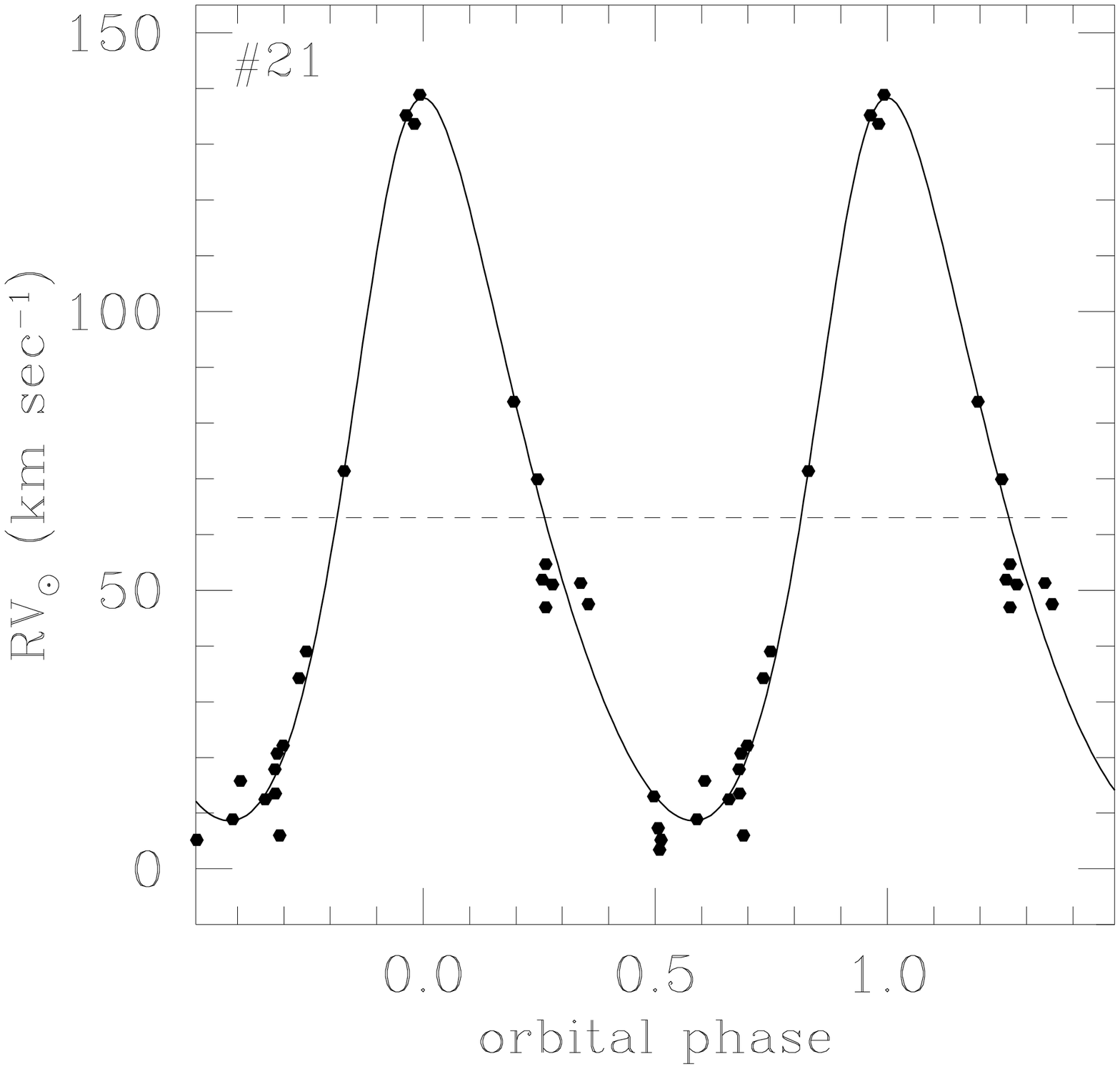,height=8cm,width=8cm}}
\caption[]{Orbital solution for star \#21. The solid line refers to the
orbital solution given in Table~4. The dashed line is the baricentric
velocity (63.0 km sec$^{-1}$).}
\end{figure}

\begin{table*}
\caption[]{Dispersion ($\sigma$), mean and error of the mean for equivalent
widths (EW) and half-intensity widths (HIW) of $\lambda$~5876~\AA\ HeI line.
The projected rotational velocities are from Equation (8).}
\begin{flushleft}
\begin{tabular}{llclllc}
\hline
\# &\multicolumn{2}{c}{EW (\AA)}&&\multicolumn{2}{c}{HIW (\AA)}
 &$V_{rot}\, {\rm sin} i$ \\ \cline{2-3} \cline{5-6}
 &\multicolumn{1}{c}{$\sigma$}&\multicolumn{1}{c}{mean}&&
 \multicolumn{1}{c}{$\sigma$}&\multicolumn{1}{c}{mean}&~~(km sec$^{-1}$)~~\\
&&&&&\\
 20~~~~ & 0.20  &   0.73$\pm$0.04 &               &0.53&3.01$\pm$0.13&   92$\pm$8  \\
 21     & 0.11  &   0.60$\pm$0.03 &               &0.62&3.14$\pm$0.14&   98$\pm$9  \\
 23     &0.010  &  0.706$\pm$0.003&~~\phantom{.}~~&0.07&1.54$\pm$0.02&   30$\pm$5  \\
\hline
\end{tabular}
\end{flushleft}
\end{table*}

\begin{figure}
\centerline{\psfig{file=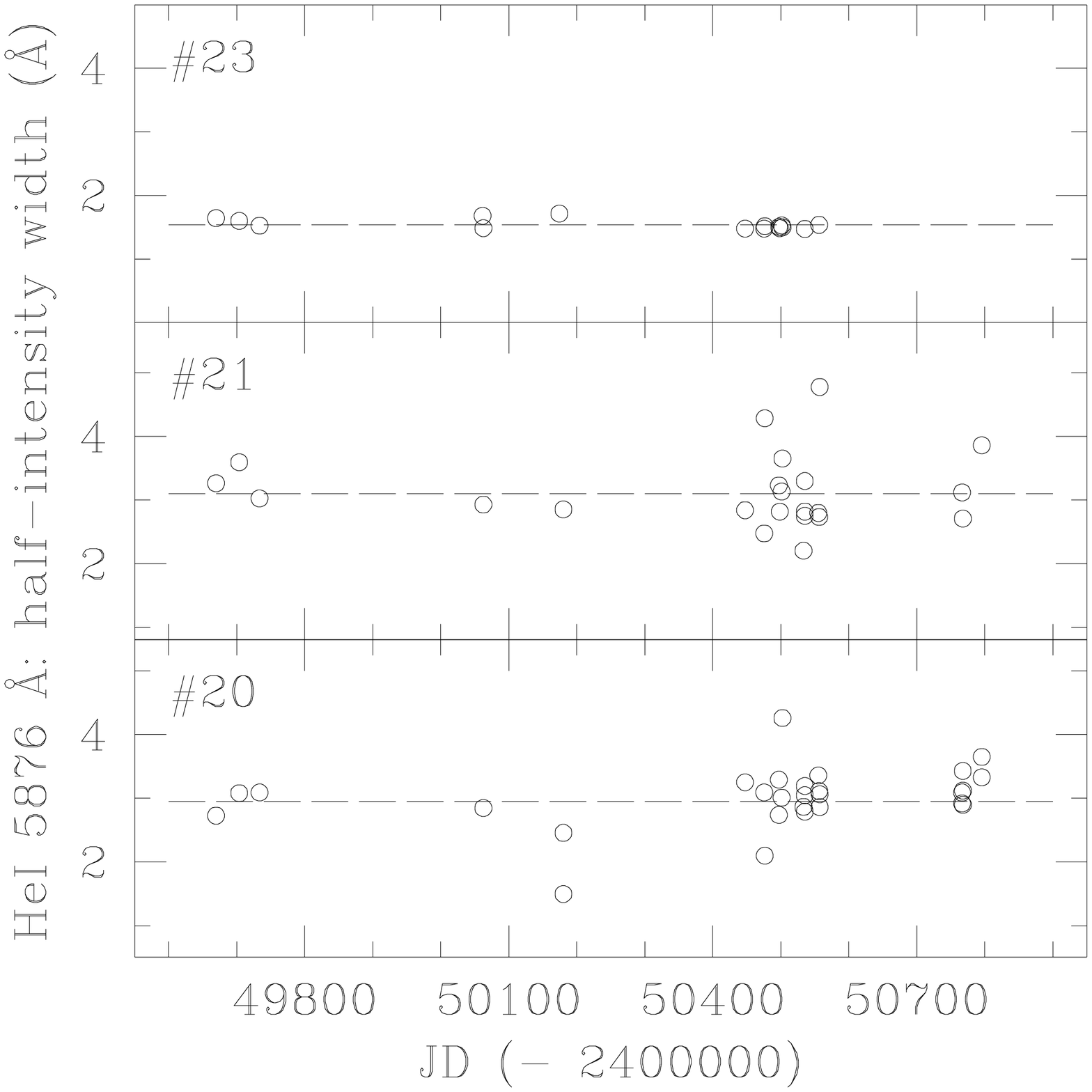,height=8cm,width=8cm}}
\caption[]{Half-intensity width (\AA) of $\lambda$ 5876 \AA\ HeI 
on spectra of programme stars \#20 (upper panel), \#21 (center) and 
\#23 (lower panel). The dashed lines represent the mean values.}
\end{figure}

\begin{figure}
\centerline{\psfig{file=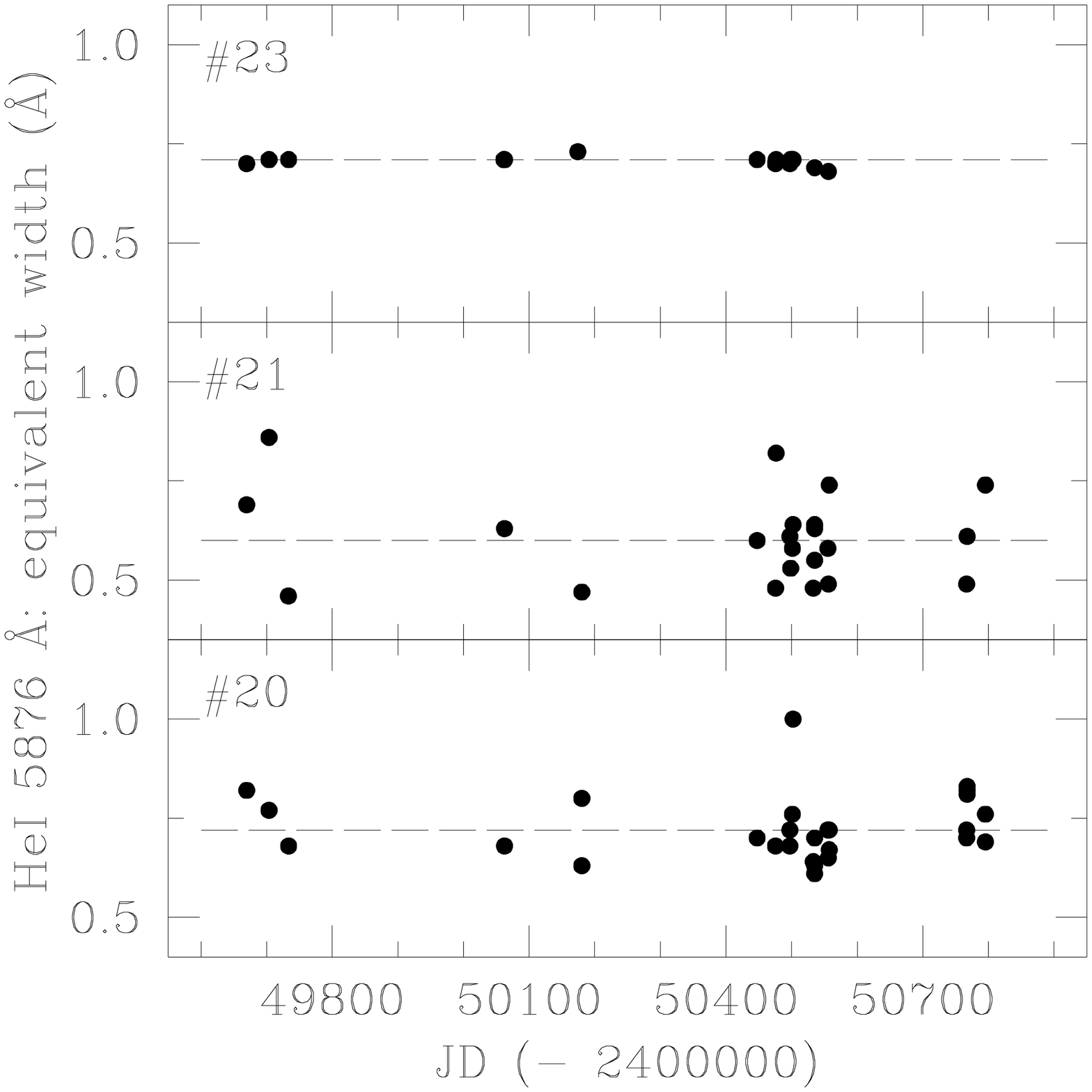,height=8cm,width=8cm}}
\caption[]{Equivalent width (\AA) of $\lambda$ 5876 \AA\ HeI 
on spectra of programme stars \#20 (upper panel), \#21 (center) and \#23 
(lower panel). The dashed lines represent the mean values.}
\end{figure}

\subsection{Spectroscopy}

Seventy-three high resolution spectra of the program stars have been secured
from Nov. 1994 to Feb. 1998 with the 1.82 m telescope and Echelle+CCD
spectrograph of the Astronomical Observatory of Padova at Asiago (Cima
Ekar). The wavelength range covered on a single frame was generally from
H$\delta$ to HeI 6678 \AA, with some spectra extending up to 9900 \AA\ and a
few others down to 3865 \AA\ (cf. Table~2 for details). The instrumental PSF
(from the FWHM of night-sky lines and thorium lines in the comparison
spectra) is generally uniform over the various order and pretty constant
during the 5 years of observations, corresponding to an average of 14 km
sec$^{-1}$ for full resolution frames and to 20 km sec$^{-1}$ for those with
a 2$\times$2 binning (corresponding respectively to 0.22 and 0.32 \AA\
resolution at H$\beta$). Details on the instrument, its performances and the
use are given in Munari \& Zwitter (1994).

The spectra have been extracted and calibrated in a standard fashion using
the IRAF reduction package. Great care has been taken to ensure the highest
quality in the wavelength calibration and therefore its suitability for
accurate radial velocity measurements. The flexure pattern of the Asiago
spectrographs (and their negligible influence on derivation of accurate RVs)
has been studied in detail by Munari \& Lattanzi (1992). The small (however
always smooth and predictable) flexures induced by telescope tracking on the
targets has been removed accordingly.

Some checks have been performed to control the accuracy of the wavelength
scale of the calibrated 1-D spectra. A brief description of the results
from these checks follows.

($i$) At the S/N ratios given in Table~2 only a few of the strongest
telluric absorption lines are detectable on a single spectrum. Therefore,
for each program star we have summed up in the velocity space all wavelength
calibrated 1-D spectra. On the summed-up spectra many telluric lines (as
weak as to produce an equivalent width of only 0.004 \AA) are measurable. 
Their mean radial velocity has turned out to be 0.0 km sec$^{-1}$ with a
dispersion of 0.52 km sec$^{-1}$, largely dominated by the difficulty to
measure so weak features. This supports the conclusion by Munari \& Lattanzi
(1992) that the Asiago Echelle+CCD spectrograph has a completely negligible
(if any) {\sl spectrograph velocity} or systematic offset.

($ii$) Our spectra have been exposed for an average of 30 minutes and at an
average height over the horizon of 38 degrees. In these condition strong
night-sky lines are expected and have been indeed recorded on our spectra.
On all spectra we have measured the three strongest night-sky lines, which
lay on separate Echelle orders. The results confirm the high control over
the wavelength calibration, with a mean radial velocity of 0.0 km sec$^{-1}$
and a typical dispersion of 0.19 km sec$^{-1}$ (the lower dispersion
compared with the weak telluric lines has to be ascribed to the higher S/N
ratio of the night-sky lines).

($iii$) Finally, nearly all spectra cover the region of the interstellar NaI
D doublet. We have measured its radial velocity, with the following results:
\begin{eqnarray}
RV_{\odot}^{NaI}(\#20)\ =\ 40.94  \pm 0.18  \ \ \ (\sigma = 0.92) \ \ \ km \ sec^{-1} \\
RV_{\odot}^{NaI}(\#21)\ =\ 41.06  \pm 0.23  \ \ \ (\sigma = 1.09) \ \ \ km \ sec^{-1} \\
RV_{\odot}^{NaI}(\#23)\ =\ 40.33  \pm 0.24  \ \ \ (\sigma = 0.95) \ \ \ km \ sec^{-1} 
\end{eqnarray}
where the standard error of the mean and the dispersion of the measurements
are given. Again, the low values for $\sigma$ confirm the accuracy and
repeatability of the wavelength scale during the night and over several
years.

\subsection{Astrometry}

Some constraints on the internal dynamics of the trapezium system
BD+00$^\circ$1617 can be derived from data available in published
astrometric catalogues, supplemented by some recent ground-based CCD
observations. BD+00$^\circ$1617 was included in the first version of the
Tycho Input Catalogue (TIC), but unfortunately it did not pass the
Recognition Process after one year of satellite operation which led to the
final version of the TIC (Halbwachs et al. 1994). Plates from the POSS-II
covering the Bochum~2 area have not yet been digitized by the time this
paper has been written, and so preliminary GSC-II data are not available.

The oldest epoch data for BD+00$^\circ$1617 come from the {\sl Astrographic
Catalogue as reduced to the ACRS} released by the United States Naval
Observatory (USNO, cf. Urban \& Corbin 1994, 1996). Bochum~2 is in the zone
photographed in 1901 by the Observatory of Algiers. Intermediate epoch data
are available from ($a$) the 1984 Palomar Schmidt plates used for the Guide
Star Catalogue (GSC-I; Lasker et al. 1990), ($b$) the astrometric reduction
of the plates of the original Palomar Sky Survey (POSS-I) performed at USNO
and known as the A1.0 Catalogue and, finally, ($c$) positions measured along
the USNO Twin Astrographic Catalogue project (Zacharias et al. 1996).

Recent epoch astrometric data come from MC95 (the X,Y in their Table~3).
Similar observations have been secured with the 1.22 m reflector of the
Asiago Astrophysical Observatory during the commissioning in Feb. 1997 of a
new CCD camera for the Newton focus (6 m focal length). For both MC95 and
Asiago data, no field stars suitable to arrange a local astrometric
reference system entered the CCD field around Bochum~2, and therefore no
reduction to J2000.0 positions has been possible. Only relative positions
between components $A$, $B$ and $C$ of BD+00$^\circ$1617 have been measured.
They are plotted in Figure~2 as the ratio of the distance from component $A$
to $B$, with respect to distance from $B$ to $C$.  The same ratio has been
computed for the other data sets above described and plotted in the figure.

A 3$\sigma$ error bar can be derived for the Astrographic Catalogue and the
CCD data, but not unambiguously for the GSC and POSS ones. The latter two
appear in Figure~2 to be of poorer astrometric quality (possibly due to the
blending with nearby faints stars of the associated cluster and the
notoriously complicated focal field geometry of Schmidt telescopes) and will
not be considered further on.

Figure~2 does not support a detectable change of the relative distances of
the three program stars ($\overline{AB} \sim \overline{BC} \sim
20.^{\prime\prime}5$) over the last century.  A least square solution (of
quite low statistical significance) may indicate a change by $\sim$0.2\%
over the time span by the data in Figure~2 (about a century). At a distance
($D$) to Bochum~2 of $\sim 6$ kpc it corresponds to a projected relative
velocity of
\begin{equation} 
V_{projected} \sim  10 \left( \frac{D}{6 \ kpc} \right) \ \ \ km \ sec^{-1} 
\end{equation} 
It does not support a runaway status for any of the program stars and it is
largely compatible inside the errors with a negative total energy (a
gravitationally bound trapezium).

In view of the limited accuracy and overall paucity of available data, a
confirmation and refinement of the astrometric results would be desiderable
through a search in plates archives for suitable old and intermediate epoch
astrometric plates covering the region of Bochum~2.

\section{Radial Velocities}

Radial velocities have been first measured on the calibrated spectra by
fitting individually each absorption line (with reference wavelengths taken
from Moore 1959). This soon led to the recognition of star \#23 as a
constant radial velocity star and the other two program stars as binaries.
The heliocentric radial velocity of star \#23 (cf. Table~3) has been found
to be:
\begin{equation}
RV_{\odot}^{\#23} = \ 68.3 \pm 0.3 \ \ \ (\sigma = 1.4) \ \ \ km \ sec^{-1}
\end{equation} 
Three times the $\sigma$ = 1.4 km sec$^{-1}$ can be taken as the threshold
for detection of binarity among the program stars. Using \#23 as a RV standard
star we have proceeded to re-evaluate radial velocities of \#20 and \#21
by cross-correlation (with the IRAF task {\sl fxcor}). The results are
listed in Table~3.

Taking a mean between the velocity of \#23 and the baricentric velocities of
\#20 and \#21 (see next section and Table~4), the trapezium (and hence
cluster) heliocentric radial velocity is (weighted mean):
\begin{equation}
RV_{\odot}^{cl} = \ 68 \pm 3 \ \ \ km \ sec^{-1}
\end{equation}
To the best of our knowledge, only Jackson et al. (1979) have so far
published a radial velocity determination for Bochum~2. They reported a 49
km sec$^{-1}$ from a few spectra of one or more of our same program stars.
Unfortunately they gave no further detail. A $\pm$7 km sec$^{-1}$ error is
attached to this value by Turbide \& Moffat (1993). Given the large
amplitude of radial velocity variation by the the program stars and the
absence of informations on the number and quality of the measurements by
Jackson et al. (1979), their RV for Bochum~2 will not be considered in the
following.

\section{Spectroscopic Orbits}

The large variability of the radial velocities of \#20 and \#21 compared to
the constancy for \#23 (cf. Table~3) argues for \#20 and \#21 to be
binaries. We have therefore proceeded with period search and determination
of the spectroscopic orbits for the latter two stars.

The orbital periods of \#20 and \#21 turned out to be $6_{.}^{d}858$ and
$11_{.}^{d}030$, respectively. The orbital solutions computed from these
orbital periods are listed in Table~4 and in graphical form are presented in
Figures~3 and 4. The errors of the orbital solutions are of the order of 5\%
or less. The larger errors for the eccentricities may results from the
uneven distribution of the observations along the orbital phase.

\begin{figure}
\centerline{\psfig{file=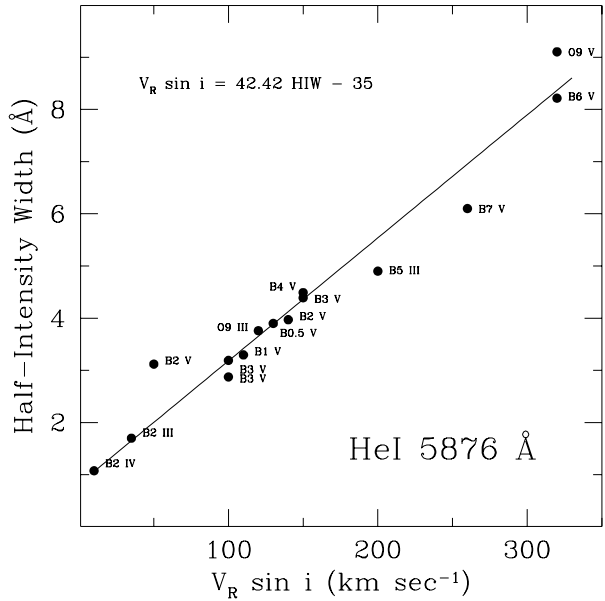,height=8cm,width=8cm}}
\caption[]{Calibration for our Echelle+CCD spectrograph of the relation
between half-intensity width (HIW) of the $\lambda$ 5876 \AA\ HeI line and
rotational velocity (standard stars from Sletteback et al. 1975). The solid
line is the least square fit given at the top of the figure (cf. Eq. 8). 
Spectral types and luminosity class of the observed stars are given next to
their symbols. HIW errors do not exceed the symbol dimension. For errors in
$V_{rot}\, {\rm sin}i$ see Sletteback et al. (1975).}
\end{figure}

\section{Rotational Velocities}

To estimate the projected rotational velocities of the program stars
($V_{rot}\, {\rm sin} i$) we have selected a sample of early type objects from
the rotational velocity standards given by Sletteback et al. (1975). We have
observed them with the same Echelle+CCD set-up used for BD+00$^\circ$1617.

Helium lines are notoriously a good choice for rotational velocity
determination in early type stars (they are less affected by Stark
broadening compared to hydrogen and are stronger than metallic lines).
$\lambda$ 5876 \AA\ is the best recorded helium line in our spectra of
BD+00$^\circ$1617 and the half-intensity width (HIW) of the $\lambda$ 5876
\AA\ HeI line has been measured on all standard stars. In Figure~7 the HIW
of the $\lambda$ 5876 \AA\ HeI line for the Sletteback et al. (1975)
standard stars is plotted against the tabulated rotational velocity. The
correction for the instrumental PSF has been applied to the data assuming
that the PSF and the rotationally broaden profile (at low $V_{rot}$) combine
as:
\begin{equation}
\sigma_{rotation}^2 = \sigma_{observed}^2 - \sigma_{PSF}^2
\end{equation}
The instrumental PSF has been derived from the profile of night-sky emission
lines (which follow the same optical path as the stellar light).  The effect
of the correction is quite small and the induced displacement in Figure~7 is
barely noticeable only for the two stars with $V_{rot}\, {\rm sin} i \ 
<$~50~km~sec$^{-1}$.

As expected, a quite straight linear relation exists in Figure~7 between
the projected rotational velocity and the HIW, equally valid for main
sequence and giant stars.  A least square fit gives:
\begin{equation}
V_{rot}\, {\rm sin} i \ =\ 42.42 \times HIW \ -\ 35 \ \ \ \ \ \ \ \  km \ sec^{-1} 
\end{equation}
The average HIW of the $\lambda$ 5876 \AA\ HeI line for the
BD+00$^\circ$1617 stars as measured on all available spectra is given in
Table~5 together with the corresponding projected rotational velocity
inferred from relation (8).

The low values of $V_{rot}\, {\rm sin} i$ for BD+00$^\circ$1617 stars agree
with the high fraction (2/3) of binaries, a general trend first noted by Abt
\& Hunter (1962 -- and confirmed by later investigations, cf. Hack \& Struve
1969) who pointed out how tidal energy dissipation in close binaries is
expected to act as a breaking mechanism on stellar rotation.

\section{Helium lines variability}

The equivalent width (EW) and half-intensity width (HIW) of the $\lambda$
5876 \AA\ HeI line have been measured on all spectra of the program stars.
The individual measures are plotted as function of time in Figures~5 and 6.
In Table~5 the dispersion of the measurements, their mean and the error of
the mean are given. There is a striking evidence: HeI lines in the two
binary stars vary by a large amount both in equivalent width and in
broadness while they are remarkably constant in the non-binary \#23 star.
The amplitude of variation is one order of magnitude the error of the
measurements (fixed by the dispersion around the mean for the non-binary
\#23) and therefore the detection has to be considered as quite robust.

The variation can in principle either be ($a$) intrinsic to the primary, or
($b$) ascribed to the visibility of the spectrum of the secondary blended
with the primary's one. The following discussion leads to the conclusion
that the variations are most probably intrinsic to the primary.

\begin{figure}
\centerline{\psfig{file=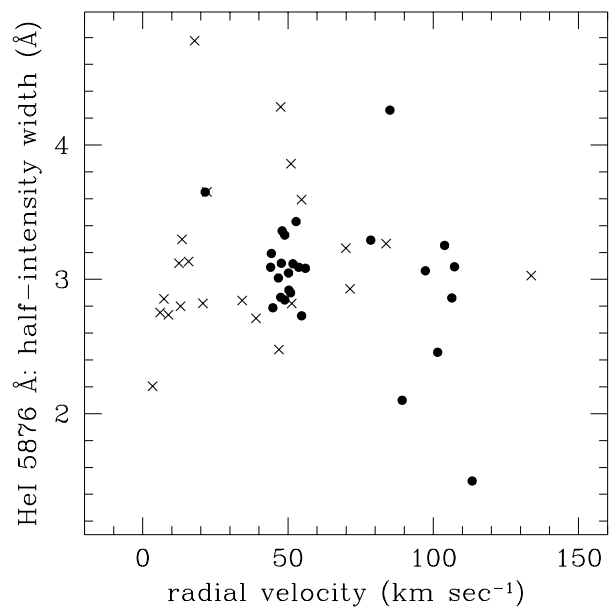,height=8cm,width=8cm}}
\caption[]{Half-intensity width (HIW) of HeI 5876 \AA\ plotted against
radial velocity for \#20 (filled circles) and \#21 (crosses). The expected
distribution for an unresolved double lined spectroscopic binary is
V-shaped, with maximum HIW reached at extrema of the velocity range. This is
clearly not the case for both stars which distributions on the graph do not
show any correlation with the radial velocity.}
\end{figure}

First, we discuss the possibility that the variations are induced by the
visibility of the spectrum of the secondary. Both the EW and HIW of the HeI
line vary by a factor of two. Thus the intensity of the HeI line in the
companion star should be similar to that of the primary and the two must
have similar brightness.  From Jasheck \& Jaschek (1987) a general upper
limit to the {\sl intrinsic} rotational velocity of late O stars may be
taken to be $V_{rot} \sim$300 km sec$^{-1}$. Comparison with the projected
rotational velocity in Table~3 gives a lower limit to the inclination of the
rotational axis. The assumption that rotational axis and the orbital axis
are roughly aligned puts the constraint on the mass of the companion from
the mass function in Table~4. For star \#20 this is
\begin{eqnarray}
i_{rot}^{\#20} \simeq i_{orb}^{\#20} \geq 18^\circ   
& \longrightarrow &  \left[ \frac{M_{2}^{3}}{(M_1 + M_2)^2} \right] _{\#20} 
\leq 0.64   \nonumber \\
& \longrightarrow & M_{2}^{\#20}\  \leq  8
 \ \ M_\odot 
\end{eqnarray}
The masses are taken as $M$(O7~V) = 28 $M_\odot$ and $M$(O9~V) = 19
$M_\odot$ (cf. Conti 1975). This {\sl maximum} mass for the secondary in
\#20 correspond to a B5 main sequence, which is vastly too dim to rival in
brightness and HeI line intensity with the O9 primary. The HeI line
variability in \#20 therefore cannot be ascribed to the visibility of the
companion and it is therefore intrinsic to the primary. 

For \#21 (an O7~V from Table~1) the situation remains unsettled 
because the same above line of reasoning on inclination leads to
\begin{eqnarray}
i_{rot}^{\#21} \simeq i_{orb}^{\#21} \geq 19^\circ   
& \longrightarrow &  \left[ \frac{M_{2}^{3}}{(M_1 + M_2)^2} \right] _{\#21} 
\leq 8.4   \nonumber \\
& \longrightarrow & M_{2}^{\#21}\ \leq  31
 \ \ M_\odot 
\end{eqnarray}
which is compatible with a companion of the same mass and therefore the
same brightness.

The most stringent argument on the nature of the HeI line variability comes
however from the search of a correlation between extrema in the HeI
variation and extrema in the radial velocities. The result is presented in
Figure~8. Maximum split between the lines of the primary and secondary (and
therefore a maximum HIW value) has to be expected when the primary is
passing through radial velocity extrema, resulting in a V-shaped
distribution of the points in Figure~8. The latter clearly shows this not
to be the case for both \#21 and \#20, which HeI line variability is
intrinsic to the O primary star and appears triggered by the binarity.

Absorption line variability is known to occur in early type stars, both
single and binary systems, although no satisfactory explanation yet exists
for the phenomenon. Jaschek, Jaschek and Kucewicz (1968) reported that in HD
125823 (B7~III) the HeI lines vary enormously from those of a B2 to those of
a B9 star, while all other lines remain constant and photometric and RV
variations are {\sl at most} marginal (Norris 1971). We too (Lattanzi,
Massone and Munari 1992) identified two helium variable stars (their mean
spectra suggesting a classification toward late main sequence B stars) in
the young open cluster NGC 225. {\sl Struve--Sahade effect} is termed by
Howarth et al. (1997) the systematic variation in relative line strength as
a function of orbital phase in double-lined spectroscopic binaries of early
types. In their investigation based on IUE spectra the frequency of
incidence of such an effect on double-lined OB spectroscopic binaries is 1
in 4. When the effect is present, radial velocity curves appear more
``noisy'' than expected solely on the base of accuracy of the measurements
(cf. Struve 1948 for the template case of the double-lined O8~V
spectroscopic binary HD 47129, the Plaskett's star).

\section{Galactic rotation}

The component of the radial velocity due to the galactic rotation may be
written as (cf. Hron 1987):
\begin{eqnarray}
RV_\odot & = & w_\circ\ {\rm sin}b \ + \ u_\circ\ {\rm cos}l\ {\rm cos}b 
               \ - \ v_\circ\ {\rm sin}l\ {\rm cos}b \nonumber \\
         &   & - \ 2[A(R - R_\circ) + \alpha (R-R_\circ)^2]\ {\rm sin}l\ {\rm cos}b 
\end{eqnarray}
where 
\begin{eqnarray}
A: \ \ Oort's \ constant &=& - \frac{R_\circ}{2}\left( \frac{d\omega}{dR}\right)_{R_\circ} \nonumber \\
\alpha : \ \ curvature \ term &=& - \frac{R_\circ}{4}\left( \frac{d^2\omega}{dR^2}\right)_{R_\circ} \nonumber \\
R &=& \sqrt{ R_{\circ}^{2} + d^2 - 2 R_\circ d {\rm cos} l} \nonumber
\end{eqnarray}
and R, R$_\circ$ are the cluster and Sun galacto-centric distances, $d$ the
distance cluster--Sun, ({\em l,b}) the heliocentric galactic coordinates of
the cluster and ($u_\circ, v_\circ, w_\circ$) is the solar motion vector.
Adopting from MC95 a cluster distance of 6 kpc and ($u_\circ, v_\circ,
w_\circ$)\ =\ (--9.32,11.18,7.0) km sec$^{-1}$ from Pont et al. (1994),
Eq.(11) rewrites as:
\begin{eqnarray}
RV_\odot & = & 13.8 \ +\ 5.8\, A\ +\ 31.8\, \alpha  
\end{eqnarray}
Hron (1987) has used distances and radial velocities to young open clusters
to investigate the rotation curve of the Galaxy. His results (valid for the
range $-3 < R - R_\circ < 5$ kpc) are best fitted by $A=17.0 \pm 1.5$ km
sec$^{-1}$ kpc$^{-1}$ and $\alpha = -2.0 \pm 0.6$ km sec$^{-1}$ kpc$^{-2}$.
It has to be noted that Bochum~2 lies at $\sim$14 kpc from the galactic
center, therefore outside the limits of applicability for Hron's relations,
but taking $\alpha = -1.4$ km sec$^{-1}$ kpc$^{-2}$ (inside the quote
errors) Eq.(12) gives a radial velocity of +68 km sec$^{-1}$, coincident
with the observed value for Bochum~2 (cf. Eq. 6).


\section*{Acknowledgments}

We would like to express our gratitude to G. Carraro and T. Zwitter who helped
to secured the spectra on Dec 16, 1994 and Jan 15, 1995, to R. Barbon and
G. Carraro for useful discussions and to R. Passuello for his expert
assistance with Linux/IRAF.


\begin{thebibliography}{}

\bibitem{} Abt H. A., Hunter J. H. jr., 1962, ApJ 136, 381
\bibitem{} Ambartsumian V. A., 1954, Comm. Byurakan Obs. 15, 3  
\bibitem{} Conti P. S., 1975, in {\sl HII regions and related topics}, T. L. Wilson
           and D. Downes ed.s, Springer-Verlag, p. 207
\bibitem{} Jackson P. D., Fitzgerlad M. P., Moffat A. F. J., 1979, in
           IAU Symp. 84 {\sl The large scale characteristics of the Galaxy}
           W. B. Burton ed., Reidel, p. 221
\bibitem{} Jaschek M., Jaschek C., Kucewicz B., 1968, Nature 225, 246
\bibitem{} Jaschek C., Jaschek M., 1987, The classification of stars,
           Cambridge Univ. Press
\bibitem{} Jeffers, H. M., van den Boss, W. H., Creeby, F. M., 1963, Index Catalogue
           of visual Double Stars, Publ. Lick Obs. 21
\bibitem{} Hack M., Struve O., 1969, Stellar Spectroscopy, Astronomical
           Observatory of Trieste
\bibitem{} Halbwachs J. L., Bessgeng G., Bastian U., Egret D., Hog E., Van
           Leeuwen F., Petersen C., Schwekendiek P., Wicenec A., 1994, A\&A 281, 25
\bibitem{} Howarth I. D., Seibert K. W., Hussain G. A. J., Prinja R. K., 1997, MNRAS 284, 265
\bibitem{} Hron J., 1987, A\&A 176, 34
\bibitem{} Lasker B. M., Sturch C. R., McLean B. J., Russell J. L., Jenkner H., Shara M. M.,
           1990, AJ 99, 2019
\bibitem{} Lattanzi M. G., Massone G., Munari U., 1991, AJ 102, 177
\bibitem{} Moffat A. F. J., Vogt N., 1975, A\&AS 20, 85
\bibitem{} Moore C. E., 1959, A Multiple Table of Astrophysical Interest, US Dept.
           of Commerce
\bibitem{} Munari U., Zwitter T., 1994, Padova and Asiago Obs. Tech. Rep. 4
\bibitem{} Munari U., Carraro G., 1995, MNRAS 277, 1269
\bibitem{} Munari U., Lattanzi M. G., 1992, PASP 104, 121
\bibitem{} Norris J., 1971, ApJS 23, 235
\bibitem{} Pont F., Mayor M., Burki G., 1994, A\&A 285, 415
\bibitem{} Salukvadze G. N., 1978, Byull. Abastumanskaya Astrofiz. Obs. 49, 39
\bibitem{} Sletteback A., Collins II G. W., Boyce P. B., White N. M., Parkinson T. D.,
           1975 ApJS 29, 137
\bibitem{} Struve O., 1948, ApJ 107, 327
\bibitem{} Turbide L., Moffat A. F. J., 1993, AJ 105, 1831
\bibitem{} Urban S. E., Corbin T. E., 1994, in {\sl Galactic and Solar System
           Optical Astrometry}, ed.  Morrison and Gilmore, Camb. Univ. Press, p.11
\bibitem{} Urban S. E., Corbin T. E., 1996, A\&A 305, 989
\bibitem{} Zacharias N., Zacharias M. I., Douglass G. G., Wycoff G. L., 1996,
           AJ 112, 2336
\end{thebibliography}
\end{document}